\documentclass[aps,twocolumn,prl,amssymb,floatfix,superscriptaddress,preprintnumbers]{revtex4}
\usepackage{graphicx,bm}

\newcommand{\SSSS}{\mbox{\boldmath ${\sf S}$} {}}

\begin{document}
\title{Evolution of Magnetic Field Generated by the Kelvin-Helmholtz Instability}

\author{M. Modestov}
\affiliation{Nordita, KTH Royal Institute of Technology and Stockholm
University, Roslagstullsbacken 23, SE-10691 Stockholm, Sweden}
\author{V. Bychkov}
\affiliation {Department of Physics, Ume{\aa} University, SE-901 87
Ume{\aa}, Sweden}
\author{G. Brodin}
\affiliation {Department of Physics, Ume{\aa} University, SE-901 87
Ume{\aa}, Sweden}
\author{M. Marklund}
\affiliation {Department of Physics, Ume{\aa} University, SE-901 87
Ume{\aa}, Sweden}
\affiliation {Department of Applied Physics, Chalmers University of
Technology, SE-41296 Gothenburg, Sweden}
\author{A. Brandenburg}
\affiliation{Nordita, KTH Royal Institute of Technology and Stockholm
University, Roslagstullsbacken 23, SE-10691 Stockholm, Sweden}
\affiliation{Department of Astronomy, Stockholm University,
SE-10691 Stockholm, Sweden}

\begin{abstract}
The Kelvin-Helmholtz instability in an ionized plasma is considered
with a focus on the generation of magnetic field via the Biermann battery
mechanisms. The problem is studied through direct numerical simulations of
two counter-directed flows in 2D geometry. The simulations demonstrate the
formation of eddies and their further interaction resulting in a large
single vortex. At early stages, the generated magnetic field evolves
due to the baroclinic term in the induction equation, revealing
significantly different structures from the vorticity field, despite the fact
that magnetic field and vorticity obey identical equations. At later times, the magnetic field
exhibits complex behavior and continues to grow even after a hydrodynamic
vortex has developed.
\end{abstract}

\preprint{NORDITA-2014-17}

\maketitle
\setlength\abovedisplayskip{6pt}
\setlength\belowdisplayskip{6pt}

\section{Introduction}

The Kelvin-Helmholtz (KH) instability is one of the most important,
fundamental and powerful phenomena in fluid mechanics and plasma physics.
The instability develops at the interface between two fluids (gases,
plasmas), when one component is gliding along the other. The most important
outcome of the KH instability in nature is the generation of turbulence
via cascades of interacting vortices. Various examples of the KH instability
may be encountered in geophysical and astrophysical flows, from ocean
surface waves exited by wind, turbulent jets and wakes, up to
large-scales instabilities in the interstellar medium, accretion discs and
supernova remnants \cite{Shen_2006,Barranco_2009,Wang_2001}. Presently,
there is also growing interest in the KH instability in the context of
inertial confinement fusion (ICF) \cite{Hurricane_2008,Hurricane_2009,Harding,Raman,Doss}.
Initially, interest in the KH instability has
been fueled by research focusing on the Rayleigh-Taylor (RT)
instability, which has been one of the most actively explored problems within the ICF
applications for decades. At the nonlinear stage of the RT instability,
light fluid (pushing or supporting a heavy one) forms bubbles rising
``up'', with spikes of heavy matter falling ``down'' in a real or effective
gravitational field \cite{Layzer,Betti_2006,Modestov_2008,Bychkov-2007}. The
relative motion of light and heavy components results in a secondary KH
instability with subsequent generation of turbulence and possible mixing of
the two substances. The well-known mushroom structure of the RT bubbles is, in
fact, the outcome of a secondary KH instability.

However, lately, a large number of papers have addressed the ICF related KH
instability for its own sake without direct relation to the RT instability
\cite{Hurricane_2008,Hurricane_2009,Harding,Raman,Doss}. A good
deal of experiments have been designed and performed on the Omega Laser
Facility focusing on the KH instability, e.g., at the foam-aluminum
interface in a layered target with two different substances set in motion by
counter-propagating shock waves \cite{Doss}. The other option was inducing
the KH instability by a shock refracted at an interface separating two
substances of noticeably different density \cite{Hurricane_2009,Harding,Raman}.
The purpose of these experiments was typically to study generation
of vortices and turbulence at the KH unstable interface. There has also been
much interest in the RT and KH hydrodynamic instabilities as sources of
magnetic field in plasmas. Several mechanisms of magnetic field generation
in plasmas have been proposed, including thermo-electric and baroclinic
effects, the ponderomotive force from an inhomogeneous laser beam, and some
others \cite{Stamper_1971,Mima_1978,Haines,Stamper_1991}.

Under extreme ICF conditions, plasma motion is expected to produce an ultra-high magnetic
field, which may alter the plasma flow dynamics as well as influence
background magnetic and electric fields. The earliest measurements of
magnetic field produced by laser plasma were made already in seventies,
detecting kilogauss field strength \cite{Stamper_1971,Askatyan}. Modern
powerful laser setups stimulate experimental activity during the last few
years in this area. Recent experiments on the RT instability at the OMEGA
laser facility demonstrated generation of magnetic field with values up to 1
MGauss \cite{Rygg_2008,Gao_2012,Manuel_2012}. In order to obtain
thorough knowledge of the KH instability phenomenon, a special setup has been
designed and built within the OMEGA facilities \cite{Hurricane_2008}.
Unfortunately, so far, experiments on the laser-driven KH instability have been
performed without direct measurements of the instability-generated magnetic
field \cite{Hurricane_2008,Hurricane_2009,Harding,Raman,Doss}.

Much work has also been done on numerical simulations of both the RT and KH
instabilities, taking into account the resulting magnetic field generation \cite{Srinivasan_PRL,Srinivasan_PoP,Modica,Alves_2012}.
However, until now, all numerical investigations have been for strongly
oversimplified model flows, such as the classical 2D RT instability at an
inert interface \cite{Srinivasan_PRL,Srinivasan_PoP,Modica}, which is
far away from realistic ICF conditions. Simulations of the
KH instability in laser plasma have also been performed, but they did not
take into account the magnetic field generation \cite{Raman}. Only
recently, magnetic field generation by the KH instability has been
considered in the astrophysical context \cite{Alves_2012}. However, those
studies have been performed within the kinetic, not magneto-hydrodynamic
(MHD) approach, for the specifically astrophysical cold-fluid KH
perturbations and electron-ion shear flows.

The purpose of the present paper is to investigate generation and evolution
of the magnetic field arising from the KH instability due to the Biermann
battery effect. Analytical treatment of the full set of MHD equations is
extremely difficult due to the nonlinear terms, although linear stability
analysis may provide necessary estimates for further experiments and
computer studies. By contrast, direct numerical simulations are a more
powerful tool, which provides a complete picture of the plasma dynamics. For
this work we have performed numerical simulations of the magnetic KH
instability using the {\sc Pencil Code} \cite{pencil_web,pencil_2002}. We
consider two counter-directed flows of conducting plasma in 2D domain which
is a natural setup for the KH studies similar to the Omega Laser facility
experiments \cite{Hurricane_2008,Hurricane_2009,Harding,Raman,Doss}.
First of all, we show that the KH instability does generate magnetic
field due to a baroclinic term in the induction equation. We observe and
discuss the dynamics of the generated magnetic field and the vorticity field in
the flow. In contrast to previous studies of the RT instability
with magnetic field generation \cite{Srinivasan_PRL,Srinivasan_PoP,Modica}, we
show that the behavior of magnetic field and vorticity in the flow may be
qualitatively different. In particular, the magnetic field may yield
complex structures influenced by secondary KH instabilities at
smaller scales. Our simulations show that the magnetic field continues to
grow even after the hydrodynamic vortex has been developed and started
decaying due to non-zero plasma viscosity. The results obtained demonstrate
that the relation between vorticity and magnetic field in the MHD
instabilities is not as straightforward, as it was believed previously, and
indicate wide prospects for future research, including both experimental,
theoretical and numerical approaches.

\section{The basic plasma model equations and the numerical method}

In order to study magnetic field generation owing to the KH instability,
we solve the compressible MHD equations for a visco-resistive plasma that
is fully ionized. The magnetic field is resolved in terms of the magnetic
vector potential $ {\bf B}=\nabla \times {\bf A}$, thus ensuring
zero divergence of ${\bf B}$. Specific entropy is used instead of temperature.
Thus, the governing equations of plasma dynamics are
\begin{equation}
\label{eq1}
\frac{D\ln \rho}{Dt}=-\nabla \cdot {\rm {\bf u}},
\end{equation}
\begin{equation}
\label{eq2}
\frac{D{\rm {\bf u}}}{Dt}=-\frac{1}{\rho }\nabla p + \frac{1}{\rho}{\rm {\bf
J}}\times {\rm {\bf B}}+\frac{1}{\rho }\nabla \cdot 2\nu \rho \SSSS,
\end{equation}
\begin{equation}
\label{eq3}
\frac{\partial {\rm {\bf A}}}{\partial t}={\rm {\bf u}}\times {\rm {\bf
B}}-\eta \mu_0 {\rm {\bf J}}+\beta \frac{1}{\rho} \nabla p,
\end{equation}
\begin{equation}
\label{eq4}
\frac{Ds}{Dt}=\frac{1}{\rho T}\nabla \cdot \left( {K \nabla T}
\right)+\frac{1}{\rho T}\eta \mu _0 {\rm {\bf J}}^2+\frac{1}{T}2\nu \SSSS^2,
\end{equation}
where $D/Dt=\partial /\partial t+{\rm {\bf u}}\cdot \nabla $ is the
advective time derivative, $\rho $ is the plasma density, ${\rm {\bf u}}$ is the
velocity, $p$ stands for the pressure, ${\rm {\bf J}}=\mu _0^{-1} \nabla \times
{\rm {\bf B}}$ is the current density, $\nu $ and $\eta $ are the kinematic
viscosity and magnetic diffusivity, respectively, $K$ is the thermal conductivity,
$\SSSS$ is the strain tensor,
\begin{equation}
\label{eq5}
{\sf S}_{ij} =\frac{1}{2}\left( {\frac{\partial u_i }{\partial x_j
}+\frac{\partial u_j }{\partial x_i }-\frac{2}{3}\delta _{ij} \nabla \cdot
{\rm {\bf u}}} \right),
\end{equation}
$\beta=m_{\rm p}/e$ is the proton mass to charge ratio, $T$ is temperature,
and $s$ is the specific entropy. The ideal gas equation complements
Eqs.~(\ref{eq1})--(\ref{eq4}), so that the pressure is given
by $p=\frac{1}{\gamma }\rho c_s^2 $, where $\gamma =c_P /c_V =5/3$ is the
ratio of specific heats at constant pressure and volume, respectively, the
sound speed is a function of density and entropy defined as $c_s^2 =c_{s0}^2
\exp \left[ {\gamma s/c_p +\left( {\gamma -1} \right)\ln \left( {\rho /\rho_0}
\right)} \right]$, and $c_{s0}$ and $\rho_0$ are normalization constants.
The last term in the induction equation (\ref{eq3}) is identical to the baroclinic
term in the vorticity equation; see Eq.~(\ref{eq10}) below. It represents the Biermann battery mechanism,
which acts as a source of magnetic field generation.

The set of equations (\ref{eq1})--(\ref{eq5}) has been
solved with the help of the {\sc Pencil Code} \cite{pencil_web,pencil_2002}, based on
sixth-order finite difference spatial derivative approximations and a
third order Runge-Kutta scheme for time stepping. The code is primarily
used to solve 3D problems, such as turbulent solar dynamo evolution in
Cartesian, spherical or cylindrical coordinates. In addition, for
efficient massive calculations the MPI parallelization can be used in all
three directions. The upper and lower walls are assumed to be
impenetrable stress-free boundaries with a perfect conductor conditions for
the magnetic vector potential, i.e.,
\begin{equation}
 \label{eq6}
 \begin{array}{l}
u_y =0, \quad \partial u_x/\partial y=0, \\
A_y =0, \quad \partial A_x/\partial y=0.
 \end{array}
\end{equation}
In the other directions we use periodic boundary conditions.
In all the simulations presented below we use $1152^2$ meshpoints.

The KH instability is essentially a 2D phenomenon, so that its main features may well be studied
in two dimensions. Taking into account 3D geometry one is faced
with turbulent mixing in the transverse direction to the initial flow plane.
This may conceal important physical properties of the instability and makes it
computationally expensive and harder to investigate.
Moreover, considering magnetic fields in
3D, there is inevitably magnetic dynamo action, which affects the
evolution of the magnetic field. In a 2D flow, magnetic fields originate
from the Biermann battery only, while the MHD dynamo (the first term of the rhs
of Eq.~(\ref{eq3})) does not operate in this case.
The Ohmic term describes the decay of magnetic field due to magnetic diffusivity.

In our studies we use a single fluid description with two layers of slightly
different density. Basically, the KH instability requires an interface with
only velocity shear, while density may be either different or the same for
both layers. The linear stability analysis for infinitesimal perturbations
$y=y_0 +f(x,t)$ with $f(x,t)\propto \exp (\sigma t+ikx)$ for the inviscid case
predicts the KH instability growth rate as \cite{Funada}
\begin{equation}
\label{eq7}
\sigma=\frac{2\sqrt \Theta }{1+\Theta}kU_0 - \frac{\Theta-1}{\Theta+1}kU_0 i,
\end{equation}
where $\Theta = \rho_{\max}/\rho_{\min}>1$ is the density ratio of the two layers, $k$ is the
perturbation wavenumber, and the plasma in the two layers moves with velocities $u_{x}=\pm U_0$.
The dispersion relation (\ref{eq7}) includes both real and imaginary parts,
although the former term is much larger than the latter one
in our simulations (see below), $\textrm{Im}[\sigma]/\textrm{Re}[\sigma]\approx 0.02$.
As we can see, the largest growth rate
corresponds to the case of equal densities of the two layers, $\Theta =1$.
The Biermann battery term generates magnetic field when the pressure and
density gradients are not collinear. In highly compressible flows, the density
gradient may arise through plasma compression. However, within this paper we
limit ourselves to an almost incompressible case; compressibility
effects will be discussed in further works. For this reason,
we use $\Theta \approx 1.02$, thus ensuring a powerful KH instability and a finite
density gradient at the interface. A higher density ratio leads to larger
gradients and a stronger effect of the Biermann battery, even though it does reduce the KH instability strength. In addition,
we solve the entropy equation, Eq.~(\ref{eq4}), which may also be considered as an
equation for temperature; it is responsible for the obliqueness between pressure
and density gradients.
For nearly isobaric flows, pressure variations arise from flow
compressibility at small but finite Mach numbers. Hence the initial
conditions for the simulations may be summarized as follows
\begin{equation}
\label{eq8}
u_x/c_{\rm s0}=0.01 \tanh ({y/d})=\ln \rho/\rho_0=-s/c_P,
\end{equation}
where $d$ is the interface thickness, which allows a continuous transition between
the two layers.
The other component of ${\rm {\bf u}}$ and all components of ${\rm {\bf B}}$
are set to zero initially.
The initial entropy distribution is computed in a way to
have constant pressure in the whole domain, so that the temperature is
inversely proportional to the density. In order to trigger the instability
we add velocity perturbations to initial distribution. We use two types if
initial perturbations as white noise and sinusoidal perturbations for $y$
component of the velocity field.

\subsection{Parameters set}

The MHD description of the KH instability involves many parameters
that may influence the evolution of the magnetic and velocity fields.
The whole parameter set in dimensionless form with the typical values
used in our simulations is listed below
\begin{equation}
\label{eq9}
 \begin{array}{l}
 {\rm Ma}=U_0 /c_s =0.01, \\
 {\rm Re}= U_0 L/\nu \approx 10^3, \\
 {\rm Re}_M = U_0 L/\eta ={\rm Re}, \\
 \rm{Pr} = \nu / \kappa =1, \\
 \Theta =1.02, \\
 \end{array}
\end{equation}
which are the Mach, Prandtl, and Reynolds numbers, the density ratio, the magnetic
Reynolds number, respectively.
Here, $\kappa=K/\rho c_P$ is the thermal diffusivity.
From a hydrodynamic point of view, the main parameters of the KH instability are
the Mach and Reynolds numbers. The Mach number quantifies the
compressibility effect; it also characterizes the time scale of the process
relative to the acoustic time scale. In
this paper we use a rather small value of the Mach number, representing an almost
incompressible flow. The Reynolds number determines also the smallest length scale
accessible and is also a limiting factor from a numerical point of view. At high
Reynolds numbers, the flow becomes turbulent. In order to
have reliable results for a turbulent flow, one has to resolve the Kolmogorov length scale, which increases dramatically the numerical resources demanded for the study.
The Prandtl number characterizes the relative role of viscous and thermal effects of the flow.
For the sake of numerical stability the Prandtl number is always set to unity
in our simulations. The density ratio of the two layers determines the
growth rate of the perturbations in the linear stage. In experiments
the actual value of the density ratio may reach several hundreds for ICF
conditions, posing an obstacle for numerical simulations.
At the same time, the density ratio was quite moderate for the KH experiments at the OMEGA laser facility being comparable to unity \cite{Hurricane_2008,Hurricane_2009,Harding,Raman,Doss}.
In this paper we also use
moderate values of this factor slightly above unity
so we can use the density for visualizing the process.

In a magnetized plasma, one additional parameter is involved in the
simulations, namely the magnetic Reynolds number.
It characterizes the decay of magnetic field due to a finite plasma conductivity.
For simplicity and the sake of numerical stability we always keep
the magnetic Reynolds number equal to the flow Reynolds number.
In addition, the hydrodynamic parameters
mentioned above may also influence the generation of magnetic field in a
critical way. For example, flow compressibility is expected to affect
the magnetic field evolution as it provides an additional contribution to
the Biermann battery term. Besides, the density gradient plays a governing
role for the magnitude of the generated magnetic field, so that the density
ratio becomes an important parameter for proper quantitative estimates.

In this paper we focus on some universal features of magnetic
field generation and its further evolution due to the KH instability. For
this reason we keep all the parameters fixed for all the simulations, using
a moderate value for the Reynolds number, ${\rm Re}\approx 10^3$, to avoid a strongly turbulent
flow. However, it is not too low either, so as to avoid fast viscous damping of
the KH instability.

\section{Results}

In the simulations we may distinguish several stages in the development
of the KH instability. In the linear stage, all perturbed values grow
exponentially -- in agreement with the dispersion relation Eq.~(\ref{eq7}). At this
stage the interface between the layers acquires a sinusoidal shape of small
amplitude, which also grows exponentially in time. During the second stage,
the amplitude continues to grow forming a number of smaller vortices, see upper panel of Fig.~1.
Later, the vortices interact with each other and merge into a single
vortex of the largest possible size allowed by the system geometry, as shown in Fig.~1.
In order to observe such interacting vortices, we have performed a simulation with $\rm{Re}=2000$;
the corresponding sequences of density and vorticity are demonstrated in Fig.~1.
After that the large-scale vortex decays due to viscosity if no external forcing
is applied to support the vorticity.
We can also observe a minor drift of the vortex core due to non-zero $\textrm{Im}[\sigma]$ in Eq.~(\ref{eq7}).
In the case of high Reynolds numbers,
the third stage may turn into turbulent mixing of the flows leading to
isotropic turbulence as the final outcome of the KH instability. In this
paper we consider the whole process of the instability development,
though paying particular attention to the early stages.
In all the simulations presented below we use a smaller Reynolds number ($\rm{Re}=1000$),
in order to avoid possible turbulent behavior and to ensure proper resolution.
\begin{figure}
\includegraphics[width=3.4in]{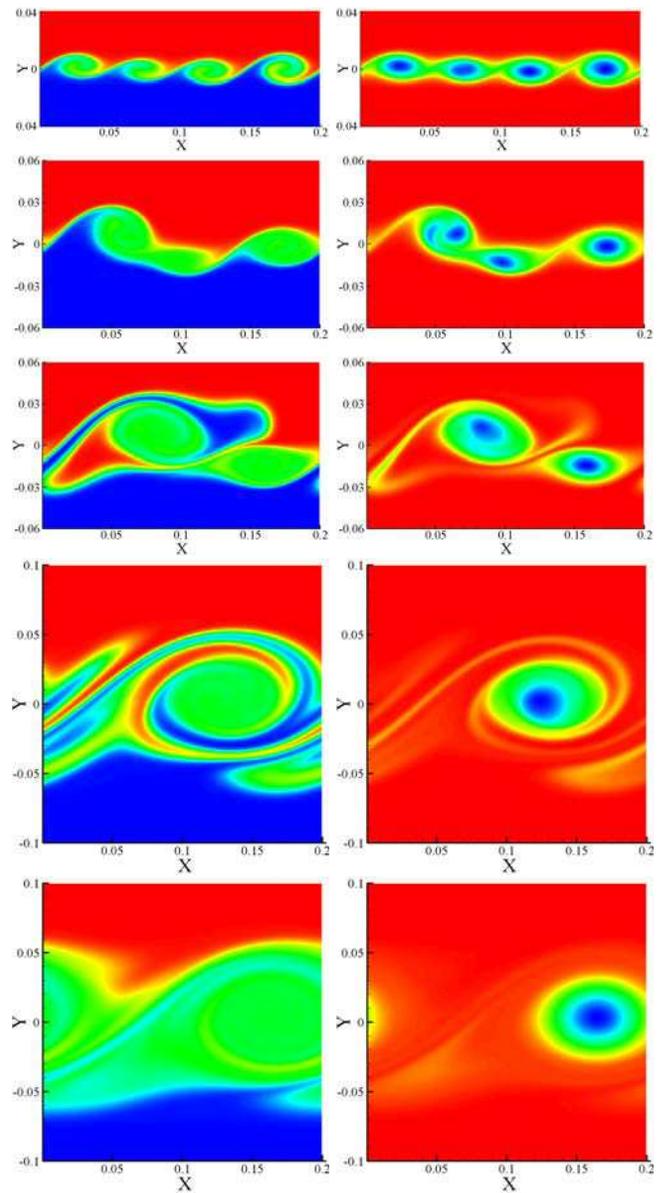}
\caption{Density (left panel) and vorticity (right panel) evolutions for the white noise simulation
with $\rm{Re}=2000$ for time instants $t=2\tau_0$; $3\tau_0$; $4.5\tau_0$; $7.5\tau_0$; $12\tau_0$.}
\end{figure}

Focusing on the magnetic field generation, we naturally expect the field to have a
similar structure as the vorticity of the flow. Analytically the evolution of
vorticity and the magnetic field are described by equations of the same
mathematic form; by taking curl of Eqs (\ref{eq2}) and (\ref{eq3}) one ends up with
\begin{equation}
\label{eq10}
\frac{\partial {\rm {\bf B}}}{\partial t}=\nabla \times ({\rm {\bf u}}\times
{\rm {\bf B}})-\beta \frac{\nabla \rho }{\rho ^2}\times \nabla p+\eta\nabla^2{\rm {\bf B}},
\end{equation}
\begin{equation}
\label{eq11}
\frac{\partial {\bm\omega} }{\partial t}=\nabla \times ({\rm {\bf u}}\times
{\bm\omega} )+\frac{\nabla \rho }{\rho ^2}\times \nabla p+\nu\nabla^2{\bm\omega}.
\end{equation}
Based on this analogy, one might be tempted to
deduce that the magnetic field is linearly proportional to the
vorticity field. In particular, such a relation between vorticity and the
magnetic field has been demonstrated in the simulations of the RT
instability in magnetized plasma \cite{Srinivasan_PoP,Modica}. By contrast,
in our simulations the generated magnetic field has significantly different
structure as compared to the vorticity, see Figs.~2--9. This difference stems
primarily from different initial conditions for these two quantities in our
simulations; these different initial conditions are expected to be
rather universal for the KH plasma experiments \cite{Hurricane_2009,
Harding,Raman,Doss}. At the initial time instant, we take zero magnetic
field everywhere in the domain, while vorticity has inevitably a certain
non-zero distribution due to the initial velocity profile forming the two
counter-flows; see in Fig.~2. As a result of this difference in the initial
conditions, the evolution of the magnetic field is mostly governed by the
second term on the right hand side of Eq.~(\ref{eq10}), while the first term can be
neglected. In the case of the vorticity equation (\ref{eq11}), the situation is the opposite
with the first term dominating over the baroclinic one.

As mentioned above, we use two types of initial perturbations, a
sinusoidal one and white noise. Below, we consider them one by one,
starting from a single mode perturbation and finishing by the white noise.
\begin{figure}
\includegraphics[width=3.4in]{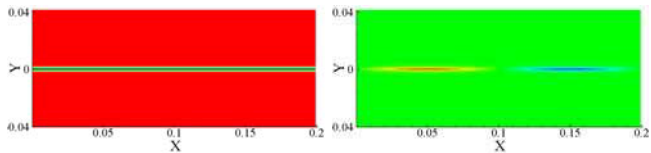}
\caption{Initial vorticity field (left) and velocity perturbation of $u_y$ (right) for a single mode simulation.}
\end{figure}
\begin{figure}
\includegraphics[width=3.4in]{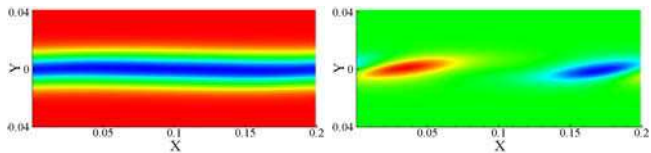}
\caption{Distribution of vorticity (left) and magnetic field $B_z$ (right) at the linear stage, $t=3\tau_0$.}
\end{figure}

\subsection{Single-mode perturbations}
A single-mode perturbation as initial condition is simple and
allows a more accurate investigation and thorough understanding of
the processes during the early stages of the instability
evolution. The initial perturbation for the transverse velocity component
represents a wave of one period with the amplitude exponentially decaying to
the outer walls,
\[
u_y=\tilde {u}_y \sin \left( {kx} \right)\exp \left( {-\left| y
\right|/w} \right),
\]
where $k=k_0=2\pi /D$, is the perturbation wavenumber, $D$ is the length of
the domain and $w$ is the interface width.
During the linear stage the instability evolution can be demonstrated by the generation of the magnetic field, as
depicted in Fig.~3. The single mode leads to the slowest growth rate of the instability, so
that after three turnover times, $t=3\tau_0$ ($\tau_0=L_0/U_0$), the
perturbations are still almost linear. The distribution of vorticity is slightly changed in terms of bending,
however it becomes much wider at the flow interface.
During the linear stage of the instability, two regions of opposite
magnetic field orientation have formed near the humps of the flow interface.
From Fig.~3 we clearly see that the generated magnetic field has quite a different structure
compared to the vorticity field, which has been discussed above.
It should be noted that the color maps for vorticity and magnetic field also differ.
Initially, the vorticity is almost zero in the hole domain (shown as red),
while in the middle region it reaches a certain negative value, depicted in dark blue.
The $z$ component of the magnetic field takes negative (blue) and positive (red) values;
this coloring is also used for all other figures.

\begin{figure}
\includegraphics[width=3.4in]{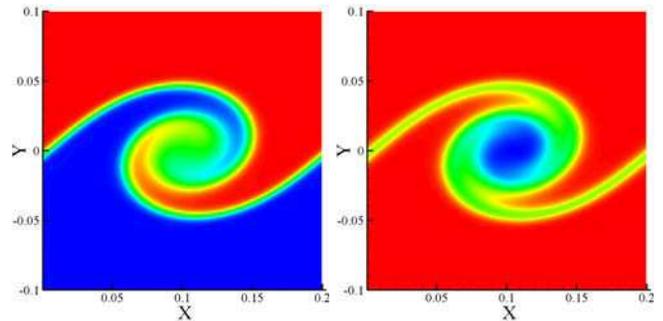}
\caption{Density (left) and vorticity (right) for $t=6\tau_0$.}
\end{figure}
\begin{figure*}
\includegraphics{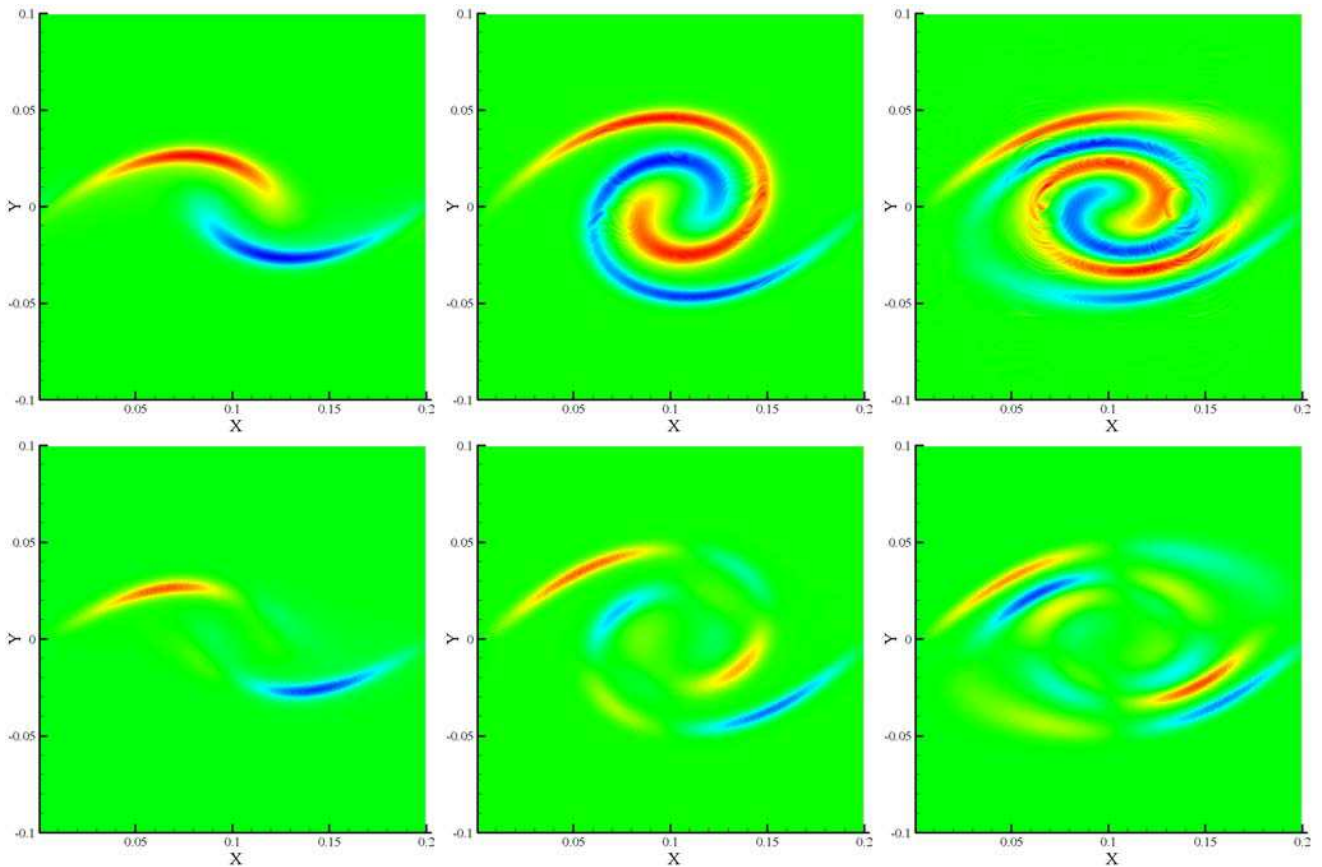}
\caption{Magnetic field structure (upper panel) and corresponding baroclinic term (lower panel) at $t=5\tau_0$; $6\tau_0$; $7\tau_0$ respectively.}
\end{figure*}
The perturbations grow further, revealing the typical picture of a wave breaking phenomenon and forming a single vortex.
In Fig.~4 we present the distribution of density and vorticity at $t=6\tau_0$.
As expected, we observe the formation of a large vortex with a characteristic structure.
During this stage the magnetic field exhibits more interesting behavior.
When the two fluids start mixing, additional regions with large density gradients
appear, leading to a spiral structure of the magnetic field.
It continues evolving mostly due to the Biermann battery term, although nonlinear effects
become more visible at this stage. We demonstrate magnetic field structure
together with the baroclinic term in Fig.~5 for several time instants.
On the left, visualizations of the magnetic field and the baroclinic term have very similar distributions.
For these time instants, the last term of Eq.~(10) is dominating over the others and governs the evolution of the magnetic field.
After a short time the structure of the baroclinic term exhibits separated islands of different signs,
following density evolution. The magnetic field tends to be continuous
due to convective term in Eq.~(10), which results in a swirling structure of the magnetic field.
However the baroclinic term remains dominating in the magnetic field generation.
In the last time instant presented in Fig.~5, we see spiral waves in the magnetic field structures.
To  make sure that the spiral waves do not stem from poor resolution, additional simulations with higher resolution have been
performed. These runs suggest that the spiral structures originate from physical effects,
e.g.\ the interference of several magnetic field sources located in different places
or a secondary instability, similar to spiral instabilities in flames \cite{Jomaas}.

Another interesting feature, which is observed in all the simulations and is in stark contrast
to the decaying vortex, is that the magnitude of the magnetic field
continues to grow during the whole process, as shown in Fig.~6. Qualitatively it may be
understood in the following way. The Biermann term evolves due to noncollinearity of density and pressure gradients.
As it is shown in Figs.~7 and 8, at late stages of the process this terms acquires a certain structure with small,
but finite magnitude, which remains almost constant. Hence, the magnetic field  has always a source
supporting its continuous growth. In Fig.~6. we also plot the time evolution of
the averaged values of the positive constituents of the baroclinic term and velocity $u_y$.
In order to present all the values in one plot, velocity and baroclinic term have been scaled to their maximal values;
the magnetic field is scaled by $\beta\langle\omega_z\rangle$,
which would be the field strength if both $B_z$ and $\omega_z$ were
solely caused by the baroclinic term.
Here, $\langle\omega_z\rangle$ is the mean vorticity at
the end of the run at $t=20\tau_0$.
Thus, if $\omega_z$ was caused by the baroclinic term,
$\langle B_z\rangle/\beta\langle\omega_z\rangle$ would be unity.
Remarkably, even though most of the contribution to $\omega_z$ comes from the
shear flow, this ratio still reaches values of about 0.1.

During the first stage, all quantities grow exponentially in a similar way.
Initially, a non-zero value of the battery term is due to sharp density gradient coupled with velocity perturbations.
As the instability develops, the velocity reaches it maximal value at $t\approx 6\tau_0 $ and starts decaying.
The baroclinic term shows similar behavior, but it decays to some small but finite level.
This value is a few orders of magnitude smaller than the initial
vorticity, so that it cannot produce additional vorticity to sustain
the vortex. It is, however, enough to sustain the growth of magnetic field, as shown in Figs.~6 and 7.
\begin{figure}[!b]
\includegraphics[trim=2cm 2cm 2cm 2.5cm, clip=true, width=3.4in]{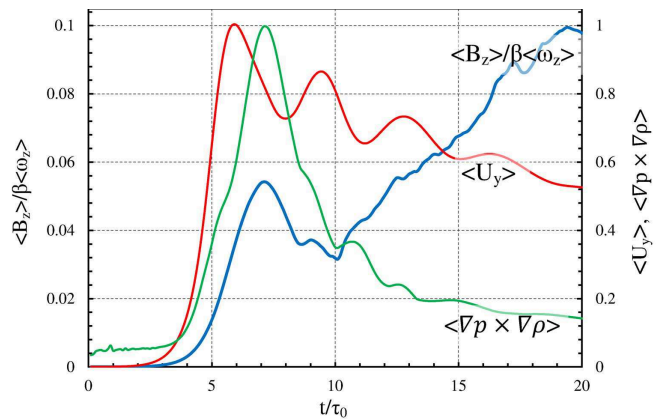}
\caption{Time evolution of the scaled averaged quantities representing magnetic field, baroclinic term and velocity $u_y$.}
\end{figure}
\begin{figure*}
\includegraphics{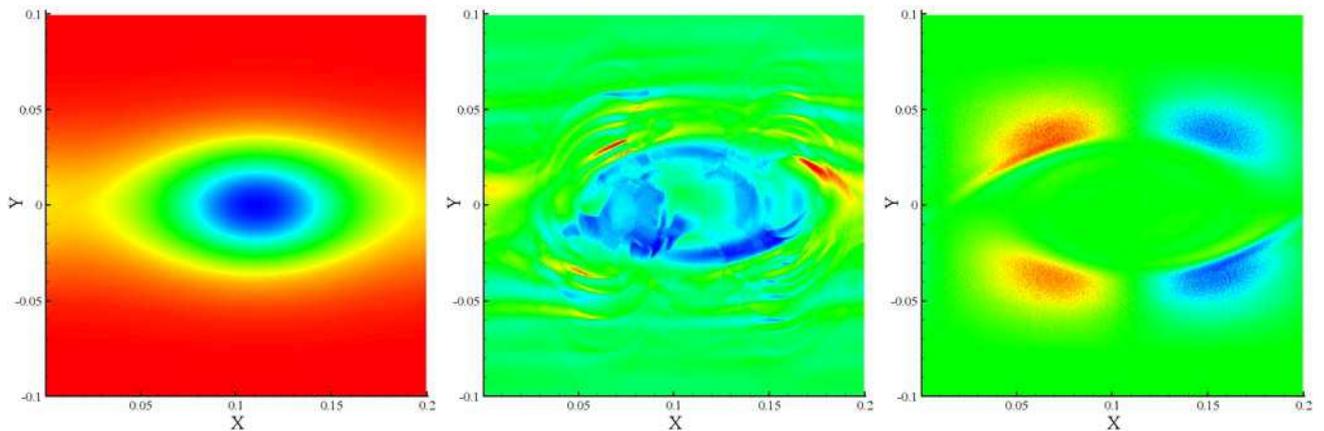}
\caption{Distribution of vorticity (left), magnetic field (middle) and baroclinic term (right) in the very later stage of the KH instability, $t=20\tau_0$.}
\end{figure*}
\begin{figure}
\includegraphics[width=3.4in]{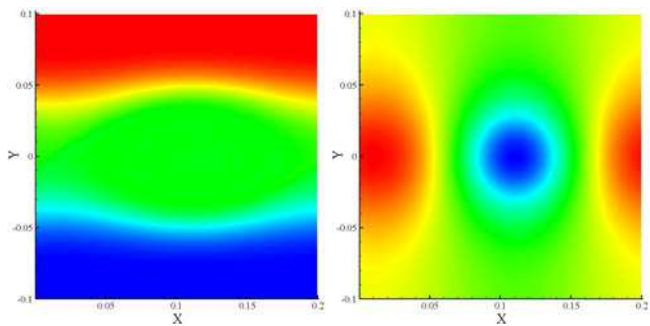}
\caption{Distributions of density (left) and pressure (right) corresponding to the previous picture.}
\end{figure}
At a much later stage, the instability demonstrates a well-developed single
vortex, which decays slowly due to viscosity; see Figs.~7 and 8.
Basically a single vortex corresponds to a circular distribution of the flow vorticity.
The elongated shape in Fig.~7 is a result of the initial flow influence,
which stretches the vortex in line with the flow. The magnetic field structure is very different
from that of both vorticity and the baroclinic term; it is an outcome of the continuous generation
due to the Biermann battery, its probable interference and convective transfer by the flow.
This phenomenon is not properly understood yet and demands further investigation well beyond the scope of the present paper.
Still we may admit that inside the vortex there is a region of negative magnetic field concentration,
which resembles the vorticity field.
The baroclinic term is a consequence of the spatial properties of flow density and pressure.
For a better understanding of its structure they are presented in Fig.~8,
which demonstrates a mixing layer with an almost homogeneous density distribution inside the vortex.
However the upper and lower parts of the domain are still filled with unmixed components,
so that there are two regions with noticeable density gradient.
The pressure picture of the vortex is governed by the hydrodynamical contribution,
as the flow remains isobaric in total. There is a certain pressure minimum in the vortex core,
due to flow compressibility; the pressure variations do not exceed the $\rm{Ma}^2$ estimate.

\begin{figure}[!b]
\includegraphics[width=3.4in]{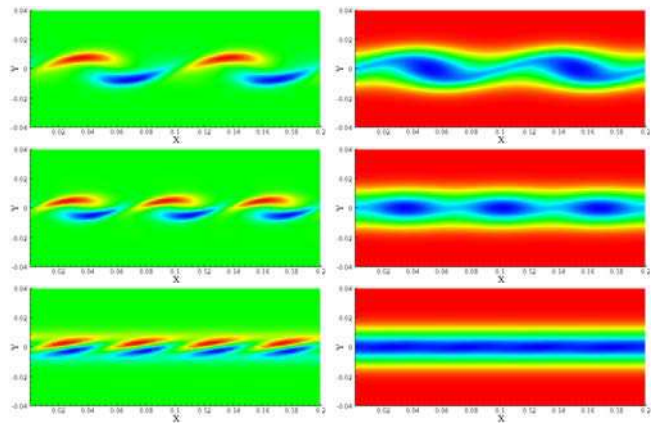}
\caption{Magnetic field (left panel) and vorticity (right panel) at time instant $t=\tau_0$ for initial perturbation of $k=2k_0$; $3k_0$; $4k_0$.}
\end{figure}
\subsection{Several-modes perturbations}
The single-mode simulations demonstrate basic features of the KH instability
together with the magnetic field generated by the plasma motion. According
to the dispersion relation (\ref{eq7}), smaller vortices with larger wavenumbers
grow faster than a single vortex of the largest possible size. In
order to study the evolution of multiple vortices in the KH instability, we have
performed several simulation runs with larger wavenumbers of initial
perturbations. In particular, we have used $k/k_0=2$, $3$, and $4$ in
order to see the development of several vortices. In
these simulations the growth of individual vortices resembles the
evolution of a single vortex described in the previous section. As expected,
the magnetic field generation occurs faster, which
is due to the larger growth rate indicated by the dispersion
relation. The early stages of vortex formation are depicted in Fig.~9 by
means of magnetic field and vorticity distributions. The vortices can be
seen both in magnetic field and vorticity, although the magnetic field represents the vortex
location more clearly, while vorticity has a very smooth profile.
\begin{figure}
\includegraphics[width=3.4in]{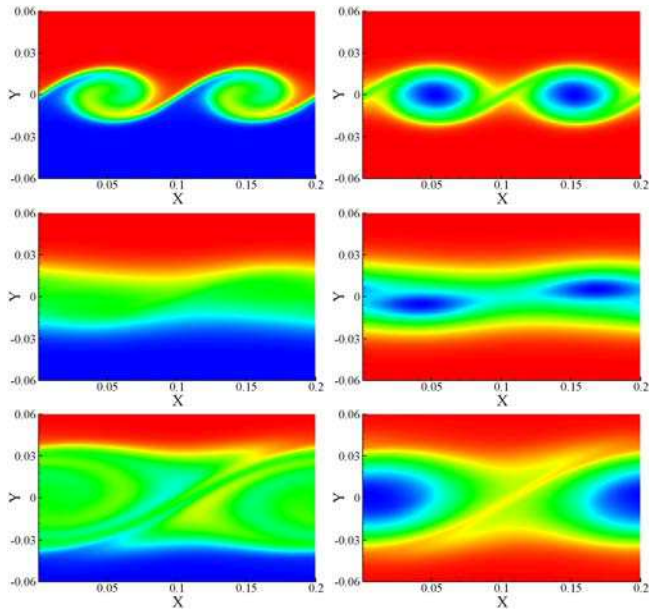}
\caption{Density (left panel) and vorticity (right panel) evolutions
during merging of two eddies for $k=2k_0$, time moments correspond to
$t/\tau_0=4$, $10$, and $15$.}
\end{figure}

After that we observe the interaction of vortices, which may be regarded as a
transient in the evolution from multiple small-scale vortices to a single large vortex
of maximal possible size. These smaller
vortices at the KH-unstable interface represent a mixing layer rather
than separately spinning eddies; this effect becomes obvious for
perturbations of high wavenumbers. Merging of vortices takes place in a
mixing layer in a smooth manner; we demonstrate it by showing the vorticity
for case $k=2k_0$ in Fig.~10. At the final stage of a single
vortex all the quantities have similar structures to those depicted in Figs.~7 and 8.
\begin{figure}[!b]
\includegraphics[width=3.4in]{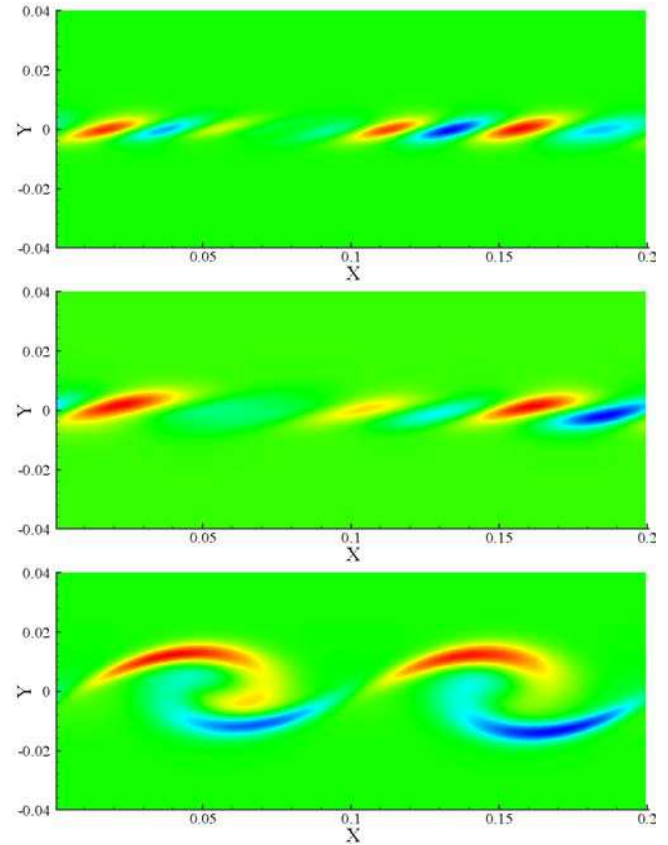}
\caption{Magnetic field evolution in the simulation with the white noise initial perturbation at time instants $t=\tau_0$ (top), $t=2\tau_0$ (middle) and $t=4\tau_0$ (bottom).}
\end{figure}

\subsection{White noise perturbation}
Finally, we consider the evolution of the magnetic KH instability with white
noise initial perturbations in velocity. Such conditions allow the system to
choose its own characteristic wavenumber.
Actually, the KH instability develops from the smallest vortices, which can be resolved by the simulation,
as its dispersion relation Eq.~(\ref{eq7}) does not imply any suppression
in the high wavenumber range \cite{Funada}.
We study the evolution of the instability by monitoring the magnetic field
at three times, $t/\tau_0=1$, 2, and 4; see Fig.~11.
At the early stage, regions of different sign of the magnetic field indicate the location and number of vortices.
We see that small vortices grow and interact with each other simultaneously until two eddies
are formed. Starting from this point, the instability evolution resembles the
case with $k=2k_0$, described in the previous section, although it grows faster.
In the double mode simulation, two eddies merge into a single vortex for about ten turnover times,
while in the white noise run the two eddies merge into a single one after about five turnover times.

The time evolution of the magnetic field is summarized in Fig.~12. It shows
the scaled average magnetic field versus time for several
types of initial perturbations. Roughly speaking, the magnetic field evolution may
be divided into two parts, where the first one corresponds to the
exponential growth during the earlier stages of the instability, and the
second part represents an almost linear growth at the later stages of the
process. In this figure with the logarithmic scale we see that the magnetic field grows faster for larger
wavenumbers of the initial perturbations according to the linear dispersion relation (\ref{eq7}).
The white noise perturbation produces weaker growth, which can be explained in the following way.
As we have seen in Fig.~11, already at $t=\tau_0$ there are several small eddies,
but an eddy is a nonlinear phenomenon. So we may conclude that the pure linear stage of the KH instability
in case of white noise perturbation is extremely short. However, the interaction of small vortices
also gives rise to exponential growth of the generated magnetic field.

The magnetic field evolution depicted in Fig.~12 has several interesting features.
For all cases, except $k=k_0$, there is a certain plateau in the growth of the magnetic field,
especially for the cases with $k\ge2k_0$. The plateau corresponds to the period when the mixing layer is formed and, hence,
the evolution of the instability slows down and the baroclinic term decreases.
In addition, each curve has one or even several pronounced peaks, e.g.\ at $t=5\tau_0$ and $10\tau_0$
for the white noise case. These peaks correspond to the smaller vortices merging into bigger ones.
Qualitatively it may be understood as an increase of the mixing of the two fluids, giving birth to additional
slices of different densities. This process leads to the origin of additional areas with
non-zero baroclinic term, which resulted in additional generation of the magnetic field.
\begin{figure}
\includegraphics[trim=2cm 2.3cm 2.5cm 2.3cm, clip=true, width=3.4in]{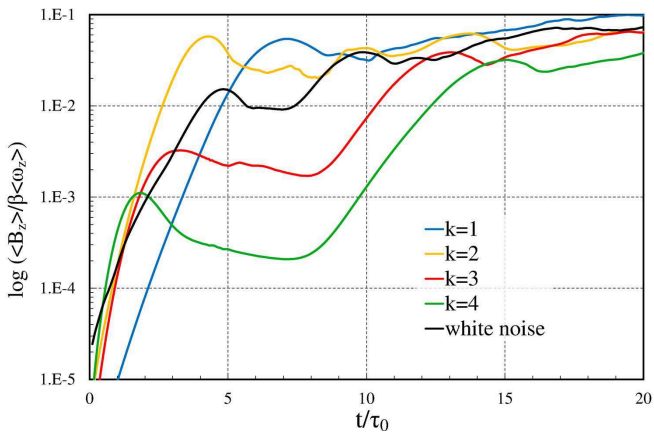}
\caption{Magnetic field evolution versus time for different initial perturbations in logarithmic scale.}
\end{figure}

\section{Conclusions}

In this paper we have investigated the KH instability in fully ionized
plasmas focusing on the generation of magnetic field through the Biermann battery.
The instability leads to growth of
magnetic field from zero, with no initial seeding. In contrast to the
related works on the RT instability with battery term \cite{Srinivasan_PoP,Modica}, we have demonstrated
that magnetic field and vorticity structures generated by the instability
may be quite different, even though they obey similar equations, Eqs.~(\ref{eq10}),
(\ref{eq11}). The important difference between magnetic field and vorticity
structures originates from intrinsically different initial conditions for
these two values, which are supposed to be the common case for the KH plasma
experiments \cite{Hurricane_2008,Hurricane_2009,Harding,Raman,Doss}.
Another important finding of the present work is that the magnetic
field continues to grow even after the largest vortex has been formed and
started decaying. It should be also mentioned that in the present
simulations, we take the flow parameters resulting in a relatively weak
generated magnetic field, so that it does not affect the hydrodynamic flow.
Our results demonstrate that the relation between vorticity and the magnetic
field in the MHD instabilities is not as straightforward, as it was believed
previously, and indicate wide prospects for future research.

\acknowledgments
We thank Dhrubaditya Mitra and Matthias Rheinhardt for help and useful comments.
Financial support from the European Research Council under the AstroDyn
Research Project 227952, the Swedish Research Council under the grants
621-2011-5076 and 2012-5797, as well as the Research Council of Norway
under the FRINATEK grant 231444 are gratefully acknowledged.


\begin{thebibliography}
\bibliographystyle{}

\bibitem{Shen_2006} Y. Shen, J. M. Stone, T. A. Gardiner, ApJ, 653, 513 (2006).
\bibitem{Barranco_2009} J. A. Barranco, ApJ, 691, 907 (2009).
\bibitem{Wang_2001} C.-Y. Wang, R. A. Chevalier, ApJ, 549, 1119 (2001).
\bibitem{Hurricane_2008} O. A. Hurricane, High Energy Density Phys. \textbf{4}, 97 (2008).
\bibitem{Hurricane_2009} O. A. Hurricane, J. F. Hansen, H. F. Robey, B. A.
Remington, M. J. Bono, E. C. Harding, R. P. Drake, and C. C. Kuranz, Phys.
Plasmas \textbf{16}, 056305 (2009).
\bibitem{Harding} E. C. Harding et al., Phys. Rev. Lett. \textbf{103}, 045005 (2009).
\bibitem{Raman} K. S. Raman, O. A. Hurricane, H.-S. Park, B. A. Remington, H.
Robey, V. A. Smalyuk, R. P. Drake, C. M. Krauland, C. C. Kuranz, J. F.
Hansen, and E. C. Harding, Phys. Plasmas \textbf{19}, 092112 (2012).
\bibitem{Doss} F. W. Doss\textit{ et al.}, Phys. Plasmas \textbf{20}, 012707 (2013).
\bibitem{Layzer} D. Layzer, Astrophys J. \textbf{122}, 1 (155).
\bibitem{Betti_2006} R. Betti and J. Sanz, Phys. Rev. Lett. \textbf{97}, 205002 (2006).
\bibitem{Modestov_2008} M. Modestov, V. Bychkov, R. Betti, and L.-E. Eriksson, Phys. Plasmas \textbf{15}, 042703 (2008).
\bibitem{Bychkov-2007} V. Bychkov, M. Modestov,  V. Akkerman, and L.-E. Eriksson, Plasma Phys. Contr. Fusion \textbf{49}, B513 (2007).
\bibitem{Stamper_1971} J. A. Stamper, K. Papadopoulos, R. N. Sudan, S. O.
Dean, E. A. McLean, and J. M. Dawson, Phys. Rev. Lett. \textbf{26}, 1012 (1971).
\bibitem{Mima_1978} K. Mima, T. Tajima, and J. N. Leboeuf, Phys. Rev. Lett. \textbf{41}, 1715 (1978).
\bibitem{Haines}M. G. Haines, Can. J. Phys. 64, 912 (1986).
\bibitem{Stamper_1991} J. A. Stamper, Laser and Particle Beams, \textbf{9}, 841 (1991).
\bibitem{Askatyan} G. A. Askat'yan, JETF Lett. \textbf{5}, 93, (1967).
\bibitem{Rygg_2008} J. R. Rygg et al, Science \textbf{319}, 1223 (2008).
\bibitem{Gao_2012} L. Gao, P. M. Nilson, I. V. Igumenschev, S. X. Hu, J. R.
Davies, C. Stoeckl, M. G. Haines, D. H. Froula, R. Betti, and D. D.
Meyerhofer, Phys. Rev. Lett. \textbf{109}, 115001 (2012).
\bibitem{Manuel_2012} M. J.-E. Manuel, C. K. Li, F. H. Seguin, J. A.
Frenje, D. T. Casey, R. D. Petrasso, S. X. Hu, R. Betti, J. Hager, D. D.
Meyerhofer, and V. Smalyuk, Phys. Plasmas \textbf{19}, 082710 (2012).
\bibitem{Srinivasan_PRL} B. Srinivasan, G. Dimonte, and X.-Z. Tang, Phys. Rev.
Lett. \textbf{108}, 165002 (2012).
\bibitem{Srinivasan_PoP} B. Srinivasan and X.-Z. Tang, Phys. Plasmas \textbf{19}, 082703 (2012).
\bibitem{Modica} F. Modica, T. Plewa, and A. Zhiglo, High Energy Density Phys. \textbf{9}, 767 (2013)
\bibitem{Alves_2012} E. P. Alves, T. Grismayer, S. F. Martins, F. Fiúza, R. A.
Fonseca, and L. O. Silva, Astrophys. J. Lett. 746:L14 (2012).
\bibitem{pencil_web} http://pencil-code.googlecode.com.
\bibitem{pencil_2002} A. Brandenburg and W. Dobler, Comput. Phys. Commun. \textbf{147}, 471 (2002).
\bibitem{Funada} T. Funada and D. D. Joseph, J. Fluid Mech \textbf{445}, 263 (2001).
\bibitem{Jomaas} G. Jomaas and C. K. Law, Phys. Fluids \textbf{22}, 124102 (2010)
\end{thebibliography}
\end{document}